\documentclass[11pt]{article}
\usepackage{epsfig,amsmath,amssymb,amsfonts,cite}
\usepackage{graphics,epsfig}
\usepackage[]{graphicx}
\usepackage{amssymb}
\usepackage{latexsym}

\begin{document}

\title{No Labeling Quantum Mechanics of Indiscernible Particles}
\author{{\sc G.
Domenech$^{1}$}\thanks{%
Fellow of the Consejo Nacional de Investigaciones Cient\'{\i}ficas y
T\'ecnicas (CONICET)}, \ {\sc F. Holik$^{1}$}, \ {\sc L.
Kniznik$^{1}$} \ and \ {\sc D. Krause$^{2}$}\thanks{Fellow of the
Conselho Nacional de Desenvolvimento Cient\'{\i}fico e Tecnol\'ogico
(CNPq)}}

\maketitle

\begin{center}

\begin{small}
${}^{1}$ Instituto de Astronom\'{\i}a y F\'{\i}sica del Espacio (IAFE)\\
Casilla de Correo 67, Sucursal 28, 1428 Buenos Aires, Argentina\\

${}^{2}$ Departamento de Filosofia - Universidade Federal de Santa Catarina,
88040-900 Florian\'opolis-SC, Brazil.
\end{small}

\end{center}

\begin{abstract}
Our aim in this paper is to show an example of the formalism we have
developed to avoid the label-tensor-product-vector-space-formalism
of quantum mechanics when dealing with indistinguishable quanta. States in this new vector space, that we call the Q-space, refer
only to occupation numbers and permutation operators act as the
identity operator on them, reflecting in the formalism the
unobservability of permutations, a goal of quasi-set theory.
\end{abstract}
\begin{small}
\centerline{\em Key words:  quasi-sets,  Fock-space, quantum
indistinguishability.}
\end{small}

%==================================================================================================

\bibliography{pom}

\newtheorem{theo}{Theorem}[section]
\newtheorem{definition}[theo]{Definition}
\newtheorem{lem}[theo]{Lemma}
\newtheorem{prop}[theo]{Proposition}
\newtheorem{axiom}[theo]{Axiom}
\newtheorem{coro}[theo]{Corolllary}
\newtheorem{exam}[theo]{Example}
\newtheorem{rema}[theo]{Remark}{\hspace*{4mm}}
\newtheorem{example}[theo]{Example}
\newcommand{\proof}{\noindent {\em Demostraci\'{o}n:
\/}{\hspace*{4mm}}}
\newcommand{\qed}{\hfill$\Box$}
\newcommand{\ninv}{\mathord{\sim}} %involutive negation
\newcommand{\Q}{$\mathfrak{Q}$}
\newcommand{\igual}{\stackrel{\tiny{\mathrm{def}}}{=}}
\newcommand{\lra}{\leftrightarrow}
\section{Introduction }

In quantum mechanics (QM), the state space of a system of
indistinguishable quanta is constructed by means of the product of
the individual spaces. The usual procedure is to first `label' the
state space of  each quanta, then making the tensor product and,
finally, imposing a symmetrization postulate \textit{by hand}. This procedure has
been largely criticized in the literature (see \cite{frekra06} for a
detailed study). To go further in this discussion, we have
explicitly constructed (based on the quasi-set theory \Q\ --see below) a formalism
within which we can obtain the same results as in the standard
formalism of quantum mechanics without appealing to artificial
labelings \cite{DHK}, \cite{laura}, \cite{frekra06}. In this paper we develop a
paradigmatic application---namely, the evaluation of the correlations
between the spin components in a two particles state---to exemplify
its use. The paper is organized as follows. We first briefly review,
in Section 2, quasi-set theory and its motivations. In Section 3 we
outline the construction of the state space in the new formalism. In
Section 4 we develop the example, and Section 5 contains the
conclusions.

\section{The basic ideas of quasi-set theory}

We briefly review here the main ideas of quasi-set theory \Q\
following mainly \cite{Unestudio}. Intuitively speaking, \Q\ is
obtained by applying ZFU-like (Zermelo-Fraenkel plus
\textit{Urelemente}) axioms to a basic domain composed of $m$-atoms
(the new ingredients that stand for indistinguishable quanta, and to which the usual concept of identity does not apply),
$M$-atoms and aggregates of them. The theory still admits a
primitive concept of quasi-cardinal, which intuitively stands for the
`quantity' of objects in a collection. This is made so that certain
quasi-sets $x$ (in particular, those whose elements are q-objects)
may have a quasi-cardinal, written $qc(x)$, but not an associated ordinal.  It
is also possible to define a translation from the language of ZFU
into the language of \Q\ in such a way so that there is a `copy' of
ZFU in \Q\ (the `classical' part of $\mathfrak Q$). In this copy,
all the usual mathematical concepts can be defined (inclusive the
concept of ordinal for the \Q-sets, the copy of standard sets in \Q), and the \Q-sets turn out to be
those quasi-sets whose transitive closure (this concept is like the
usual one) does not contain $m$-atoms.

In \Q, `pure' quasi-sets have only $m$-atoms as elements (although
these elements may be not always indistinguishable from one
another), and to them it is assumed that the usual notion of
identity cannot be applied (the expressions $x=y$ and its
negation, $x \not= y$, are not well formed formulas if either $x$ or
$y$ stand for $m$-atoms). Notwithstanding, there is a primitive
relation $\equiv$ of indistinguishability having the properties of
an equivalence relation, and a defined concept of \textit{extensional
identity}, not holding among $m$-atoms, which has the
properties of standard identity of classical set theories. More
precisely, we write $x =_E y$ ($x$ and $y$ are extensionally
identical) iff they are both qsets having the same elements (that
is, $\forall z (z \in x \lra z \in y)$) or they are both $M$-atoms
and belong to the same qsets (that is, $\forall z (x \in z \lra y
\in z)$). From now on, we shall use the symbol ``='' for the
extensional equality, except when explicitly mentioned.

Since the elements of a quasi-set may have properties (and satisfy
certain formulas), they can be regarded as
\textit{indistinguishable} without turning to be \textit{identical}
(that is, being \textit{the same} object), that is, $x \equiv y$
does not entail $x=y$. Since the relation of equality (and the
concept of identity) does not apply to $m$-atoms, they can also be
thought of as entities devoid of individuality. For details about
\Q\ and about its historical motivations, see \cite[Chap.\
7]{frekra06}.

One of the main features of \Q\ is its ability to take into account
in `set-theoretical terms' the non observability of permutations in quantum physics, which is
one of the most basic facts regarding indistinguishable quanta. In
standard set theories, if $w \in x$, then of course $(x - \{w \})
\cup \{z\} = x$ iff $z = w$. That is, we can `exchange' (without
modifying the original arrangement) two elements iff they are
\textit{the same} element, by force of the axiom of extensionality.
But in \Q\ there is a theorem guarantying the unobservability of permutations; in other words,

\begin{theo}
Let $x$ be a finite quasi-set such that $x$ does not contain all
indistinguishable from $z$, where $z$ is an $m$-atom such that $z
\in x$. If $w \equiv z$ and $w \notin x$, then there exists $w'$
such that $(x - z') \cup w' \equiv x$
\end{theo}
Here $z'$ and $w'$ stand for a quasi-set with quasi-cardinal 1 whose
only element is indistinguishable (but not identical) from $z$ and
$w$ respectively.

\section{The state space $\mathbb{V}_{Q}$}\label{eQ}

In QM, the mathematical interpretation of a physical system is a
complex separable Hilbert space ${\mathcal{H}}$. Quantum systems of
distinguishable many particles are mathematically represented by the
tensor products of their individual Hilbert spaces \cite{product},
the so called `labeled-tensor-product' of state spaces. To pick up
physical states from the whole set allowed by the labeled product, a
symmetrization postulate is added that restricts the states
available for the particles by imposing that, if particles are
indistinguishable, then they can only access symmetrized (with
respect to particle interchange) states if they are bosons, or
antisymmetrized states if they are fermions. This trick of first
labeling and then restricting the available states allows to
reproduce quantum statistics satisfactorily, without dropping
particle indexation. But this procedure may be criticized, in
particular for the necessity of first labeling the state spaces of
the indistinguishable quanta and later masking this labeling with
the addition of a postulate. Sometimes  it is argued that the usual
procedure to construct the set of possible states should be replaced
by another one that does not index particles, for example the
Fock-space formalism. However the formal construction of the
Fock-space does use the standard set theoretical framework, which
presupposes classical individuality on its foundations. This seems
to be, thus, not a genuine solution \cite{kra08}. To solve it, we have proposed
an alternative procedure (\cite{DHK}, \cite{laura}) that resembles
that of the Fock-space formalism but based on \Q, thus genuinely
avoiding artificial labeling.

Let us recall that the usual procedure in the Fock-space formalism
 is to define the fundamental (vacuum) state
$|0\rangle$ as the eigenvector of the particle number operator
$N_{k}=a_{k}^{\dag}a_{k}$ with $0$ eigenvalue, being
$a_{k}^{\dagger}$ and $a_{k}$ the creation and annihilation
operators of particles of $k$-type respectively, which satisfy
commutation relations for bosons and anticommutation relations for
fermions. The complete basis for $\mathcal{H}$ is obtained by
successive application of $a^{\dagger}$ to the vacuum state. All
operators and wave functions may be written in terms of
$a_{k}^{\dagger}$ and $a_{k}$ \cite{Robertson}.

We outline now the construction of the state space $\mathbb{V}_{Q}$
that respects indistinguishability in all steps, mainly following
\cite{DHK}. We shall be working within \Q. Let us consider a quasi-set
$\epsilon= \{\epsilon_{i}\}_{i \in I }$, where $I$ is an arbitrary
collection of indexes (this makes sense in the `classical part' of \Q). We take the elements $\epsilon_{i}$ to
represent the eigenvalues of a physical magnitude of interest.
Consider then the quasi-functions $f$ (this concept generalizes that of function), $f:\epsilon \longrightarrow
\mathcal{F}_{p}$, where $\mathcal{F}_{p}$ is the quasi-set formed of
finite and pure quasi-sets. $f$ is the quasi-set formed of
ordered pairs $\langle \epsilon_{i};x\rangle$ with
$\epsilon_{i}\in\epsilon$ and $x\in\mathcal{F}_{p}$. Let us choice
these quasi-functions in such a way that whenever $\langle
\epsilon_{i_{k}};x\rangle$ and $\langle \epsilon_{i_{k'}};y\rangle$
belong to $f$ and $k\neq  k'$, then $x\cap y=\emptyset$. Let us
further assume  that the sum of the quasi-cardinals of the
quasi-sets which appear in the image of each of these
quasi-functions is finite, and then, $qc(x)=0$ for every $x$ in the
image of $f$, except for a finite number of elements of $\epsilon$.
Let us call $\mathcal{F}$ the quasi-set formed of these
quasi-functions. If $\langle x;\epsilon_{i}\rangle$ is a pair of
$f\in\mathcal{F}$, we will interpret that the energy level
$\epsilon_{i}$ has occupation number $qc(x)$. These quasi-functions
will be represented by symbols such as
$f_{\epsilon_{i_{1}}\epsilon_{i_{2}}\ldots\epsilon_{i_{m}}}$ (or by
the same symbol with permuted indexes). This indicates that the
levels $\epsilon_{i_{1}}\epsilon_{i_{2}}\ldots\epsilon_{i_{m}}$ are
occupied. It will be taken as convention that if the symbol
$\epsilon_{i_{k}}$ appears $j$-times, then the level
$\epsilon_{i_{k}}$ has occupation number $j$. The levels that do not
appear have occupation number zero.

It is important to point out that the order of the indexes in
$f_{\epsilon_{i_{1}}\epsilon_{i_{2}}\ldots\epsilon_{i_{n}}}$ has no
meaning at all because up to now, there is no need to define any
particular order in $\epsilon$, the domain of the quasi-functions of
$\mathcal{F}$. Nevertheless, we may define an order in the following
way. For each quasi-function $f\in\mathcal{F}$, let
$\{\epsilon_{i_{1}}\epsilon_{i_{2}}\ldots\epsilon_{i_{m}}\}$ be the
quasi-set formed by the elements of $\epsilon$ such that
$\langle\epsilon_{i_{k}},X\rangle\in f$ and $qc(X)\neq  0$ ($k=
1\ldots m$). We call $supp(f)$ this quasi-set (the \textit{support}
of $f$). Then consider the pair $\langle o,f\rangle$, where $o$ is a
bijective quasi-function
$o:\{\epsilon_{i_{1}}\epsilon_{i_{2}}\ldots\epsilon_{i_{m}}\}\longrightarrow
\{1,2,\ldots,m\}.$ Each of these quasi-functions $o$ define an order
on $supp(f)$. For each $f\in\mathcal{F}$, if $qc(supp(f))= m$, then,
there are $m!$ orderings. Then, let $\mathcal{O}\mathcal{F}$ be the
quasi-set formed by all the pairs $\langle o,f\rangle$, where
$f\in\mathcal{F}$ and $o$ is a a particular ordering in $supp(f)$.
Thus, $\mathcal{O}\mathcal{F}$ is the quasi-set formed by all the
quasi-functions of $\mathcal{F}$ with ordered support. For this
reason,
$f_{\epsilon_{i_{1}}\epsilon_{i_{2}}\ldots\epsilon_{i_{n}}}\in
\mathcal{O}\mathcal{F}$ refers to a quasifunction $f\in\mathcal{F}$
and a special ordering of
$\{{\epsilon_{i_{1}}\epsilon_{i_{2}}\ldots\epsilon_{i_{n}}}\}$. For
the sake of simplicity, we will use the same notation as before. But
now the order of the indexes {\it is meaningful}. It is also
important to remark, that the order on the indexes must not be
understood as a labeling of particles, for it easy to check, as
above, that the permutation of particles does not give place to a new
element of $\mathcal{O}\mathcal{F}$. This is so because a
permutation of particles operating on a pair $\langle
o,f\rangle\in\mathcal{O}\mathcal{F}$ will not change $f$, and so,
will not alter the ordering. We will use the elements of
$\mathcal{O}\mathcal{F}$ later, when we deal with fermions.

A linear space structure is required to adequately represent quantum
states. To equip $\mathcal{F}$ and $\mathcal{OF}$ with such a
structure, we need to define two operations ``$\star$" and ``$+$", a
product by scalars  and an addition of their elements, respectively.
Call $C$ the collection of quasi-functions which assign to every
$f\in \mathcal{F}$ (or $f\in \mathcal{O}\mathcal{F}$) a complex
number (again, built in the `classical part' of \Q). That is, a quasi-function $c\in C$ is a collection of
ordered pairs $\langle f;\lambda\rangle$, where $f\in \mathcal{F}$
(or $f\in \mathcal{O}\mathcal{F}$) and $\lambda$ a complex number.
Let $C_{0}$ be the subset of $C$ such that, if $c\in C_0$, then
$c(f)=0$ for almost every $f\in \mathcal{O}\mathcal{F}$ (i.e.,
$c(f)=0$ for every $f\in \mathcal{O}\mathcal{F}$ except for a finite
number of quasi-functions). We can define in $C_{0}$ a sum and a
product by scalars in the same way as it is usually done with
functions as follows:
\begin{definition}
Let $\alpha$, $\beta$ and $\gamma$ $\in \mathcal{C}$, and $c$,
$c_{1}$ and $c_{2}$ be  quasi-functions of $C_{0}$, then
$$(\gamma\ast c)(f) := \gamma(c(f))\ \ and \ \
(c_{1}+c_{2})(f) :=  c_{1}(f) + c_{2}(f)$$
\end{definition}
\noindent The quasi-function $c_{0}\in C_{0}$ such that $c_{0}(f)=
0$, for any $f\in F$, acts as the null element of the sum, for $
(c_{0}+c)(f)= c_{0}(f)+c(f)= 0+c(f)= c(f), \forall f.$ With the sum
and the multiplication by scalars defined above we have that
$(C_{0},+,\ast)$ is a complex vector space. Each one of the
quasi-functions of $C_{0}$ should be interpreted in the following
way: if $c\in C_{0}$ (and $c\neq c_{0}$), let $f_{1}$, $f_{2}$,
$f_{3}$,$\ldots$, $f_{n}$ be the only functions of $C_{0}$ which
satisfy $c(f_{i})\neq  0$ ($i= 1,\ldots,n$). These quasi-functions
exist because, as we have said above, the quasi-functions of $C_{0}$
are zero except for a finite number of quasi-functions of
$\mathcal{F}$. If $\lambda_{i}$ are complex numbers which satisfy
that $c(f_{i})= \lambda_{i}$ ($i= 1,\ldots,n$), we will make the
association
$c\approx(\lambda_{1}f_{1}+\lambda_{2}f_{2}+\cdots+\lambda_{n}f_{n})$.
The symbol $\approx$ must be understood in the sense that we use
this notation to represent the quasi-function $c$. The idea is that
the quasi-function $c$ represents the pure state which is a linear
combination of the states represented by the quasi-functions $f_{i}$
according to the interpretation given above.

In order to calculate probabilities and mean values, we have to
introduce a scalar product, in fact two of them:  $\circ$ for bosons
and $\bullet$ for fermions, thus obtaining two (normed) vector
spaces $({\mathbb{V}_{Q}},\ \circ )$ and  $({\mathbb{V}_{Q}},\
\bullet )$ :
\begin{definition}
Let $\delta_{ij}$ be the Kronecker symbol and
$f_{\epsilon_{i_{1}}\epsilon_{i_{2}}\ldots\epsilon_{i_{n}}}$ and
$f_{\epsilon_{i'_{1}}\epsilon_{i'_{2}}\ldots\epsilon_{i'_{m}}}$ two
basis vectors, then
$$
f_{\epsilon_{i_{1}}\epsilon_{i_{2}}\ldots\epsilon_{i_{n}}}\circ
f_{\epsilon_{i'_{1}}\epsilon_{i'_{2}}\ldots\epsilon_{i'_{m}}} :=
\delta_{nm}\sum_{p}\delta_{i_{1}pi'_{1}}\delta_{i_{2}pi'_{2}}\ldots\delta_{i_{n}pi'_{n}}
$$
The sum is extended over all the permutations of the set
$i'=(i'_{1},i'_{2},\ldots,i'_{n})$ and for each permutation $p$,
$pi'=(pi'_{1},pi'_{2},\ldots,pi'_{n})$.
\end{definition}
This product can be easily extended over linear combinations.
\begin{definition}
Let $\delta_{ij}$ be the Kronecker symbol,
$f_{\epsilon_{i_{1}}\epsilon_{i_{2}}\ldots\epsilon_{i_{n}}}$ and
$f_{\epsilon_{i'_{1}}\epsilon_{i'_{2}}\ldots\epsilon_{i'_{m}}}$ two
basis vectors, then
$$f_{\epsilon_{i_{1}}\epsilon_{i_{2}}\ldots\epsilon_{i_{n}}}\bullet
f_{\epsilon_{i'_{1}}\epsilon_{i'_{2}}\ldots\epsilon_{i'_{m}}} :=
\delta_{nm}\sum_{p}\sigma_{p}\delta_{i_{1}pi'_{1}}\delta_{i_{2}pi'_{2}}\ldots\delta_{i_{n}pi'_{n}}$$
where: $s^{p}=+1$ if $p$ is even and $s^{p}= -1$ if $p$ is odd.
\end{definition}
The result of this second product $\bullet$ is an antisymmetric sum
of the indexes which appear in the quasi-functions. In order that
the product is well defined, the quasi-functions must belong to
$\mathcal{O}\mathcal{F}$. Once this product is defined over the
basis functions, we can extend it to linear combinations, in a
similar way as for bosons. If the occupation number of a product is
more or equal than two, then the vector has null norm. As it is a
vector of null norm, the product of this vector with any other
vector of the space would yield zero, and thus the probability of
observing a system in a state like it vanishes. This means that we
can add to any physical state an arbitrary linear combination of
null norm vectors for they do not contribute to the scalar product,
which is the meaningful quantity.

With these tools and using the language of \Q, the formalism of QM
may be completely rewritten giving a straightforward answer to the
problem of giving a formulation of QM in which intrinsical
indistinguishability is taken into account from the beginning,
without artificially introducing extra postulates. We make the
following association in order to turn the notation similar to that
of the standard formalism. For each quasi-function
$f_{\epsilon_{i_{1}}\epsilon_{i_{2}}\ldots \epsilon_{i_{n}}}$ of the
quasi-sets $\mathcal{F}$ or $\mathcal{O}\mathcal{F}$ constructed
above, we will write $\alpha
f_{\epsilon_{i_{1}}\epsilon_{i_{2}}\ldots\epsilon_{i_{n}}}:=
\alpha|\epsilon_{i_{1}}\epsilon_{i_{2}}\ldots \epsilon_{i_{n}})$
 with the obvious corresponding generalization for linear
combinations.  Once normalized to unity, the states constructed
using \Q, are equivalent to the symmetrized vectors for bosonic
states and we have shown that commutation relations equivalent to
the usual ones hold, thus being both formulations equivalent for
bosons.

For fermions, there are some subtleties involved in the
construction. First of all, let us recall the action of the creation
operator $c^{\dag}_\alpha$: let $\zeta$ represent a collection of
indexes with non null occupation number, then $
C^{\dag}_\alpha|\zeta)=|\alpha\zeta)$. If $\alpha$ was already in
the collection $\zeta$, then $|\alpha\zeta)$ is a vector with null
norm. As said above, to have null norm implies that
$(\psi|\alpha\zeta)=0$ for all $|\psi)$. Moreover, if a linear
combination of null norm vectors were added to the vector
representing the state of a system, this addition would not give
place to observable results because the terms of null norm do not
contribute to the mean values or to the probabilities. In order to
express this situation, we define the following relation:
\begin{definition}\label{e:chirimbolo} Two vectors $|\varphi)$ and $|\psi)$
are similar (and we will write $ |\varphi)\cong|\psi)$)) if the
difference between them is a linear combination of null norm
vectors.
\end{definition}
With all of this, it is straightforward to demonstrate the
equivalence of the anticommutation relations in $\mathbb{V}_{Q}$ and
in the standard Fock-space. Thus, we can conclude that both
formulations are equivalent also for fermions.

To avoid particle labeling in the expressions for observables, in
Fock-space formalism they are written in terms of creation and
annihilation operators. This is also the case in $\mathbb{V}_{Q}$.
For example, we have shown that operators $T$ acting over a single
particle states are of the form:
\begin{equation}\label{e:F}
T=\sum_{\alpha\beta}t_{\alpha\beta}a^{\dag}_{\alpha}a_{\beta}=
\sum_{k}(\alpha|k)t_{k}(k|\beta)a^{\dag}_{\alpha}a_{\beta}=\\
\sum_{k}\sum_{j}(\alpha|k)(k|T(1)|j)(j|\beta)a^{\dag}_{\alpha}a_{\beta}
\end{equation}
Interaction operators act over spaces of a greater number of
particles. The expression of an interaction operator $V$ between two
particles is:
\begin{equation}\label{e:bosdoscuerpos}
V=\frac{1}{2}\sum_{\alpha}\sum_{\beta}\sum_{\gamma}\sum_{\delta}V_{\alpha\beta,\gamma\delta}
a^{\dag}_{\alpha}a^{\dag}_{\beta}a_{\gamma}a_{\delta}=\\
\frac{1}{4}\sum_{\alpha}\sum_{\beta}\sum_{\gamma}\sum_{\delta}(\kappa\lambda|V|\mu\nu)
a^{\dag}_{\alpha}a^{\dag}_{\beta}a_{\gamma}a_{\delta}
\end{equation}

\section{Correlation in a two-particle state}
In this section we show an application of the use of the formalism
in $\mathbb{V}_{Q}$ to illustrate how the usual results of standard
QM formalism are obtained. To do so, let us consider a two spin
$1/2$ quanta regarding only to spin degrees of freedom. Let
$S_{i}=({\hbar}/2)\sigma_{i}$ be the spin operator, $\sigma_{i}$ the
Pauli matrices. We use eq. (\ref{e:F}) to write $\sigma_{z}$:
$\sigma_{z}=\sum_{\alpha\beta}(\sigma_{z})_{\alpha\beta}\
C^{\dag}_{\alpha}C_{\beta}=C^{\dag}_{+}C_{+}-C^{\dag}_{-}C_{-}$. To
obtain the spin operator in an arbitrary direction
$\hat{n}=(\sin\theta\cos\phi,\sin\theta\sin\phi,\cos\theta)$, we
propose that $\sigma_{n}$ is of the form: $$\sigma_{n}=\cos\theta\
C^{\dag}_{+}C_{+}+e^{-i\phi}\sin\theta
C^{\dag}_{+}C_{-}+e^{i\phi}\sin\theta C^{\dag}_{-}C_{+}-\cos\theta\
C^{\dag}_{-}C_{-}$$ In fact, this operator rotates the basis vectors
as usual. Thus, the mean value of $\sigma_{n}$ in the one particle
state `up' in direction $\hat{z}$ results:
$$(+|\sigma_{n}|+)=\cos\theta(+|+)+e^{i\phi}\sin\theta
(+|-)=\cos\theta$$

Now we consider the a pair of indistinguishable fermions, one with
spin `up' and the other with spin `down'. In $\mathbb{V}_{Q}$ its
state is $|+-)$. It has not to be confused with the standard
$|+-\rangle$, which is not an antisymmetric state. We first show
that in the same spatial direction, say $\hat{z}$, the spin
components are in perfect anticorrelation. As usual, the correlation
is evaluated as the mean value of an operator that represents the
measurement of $\sigma_{z}$ for both components over the state.
Differently from the standard formulation, where this operator is
obtained in the labeled tensor product space, here it is obtained
from eq. (\ref{e:bosdoscuerpos}) for the fermionic case:
\begin{equation}\label{e:zz}
\begin{split}
\mathbf{\sigma_{zz}}&=\frac{1}{2}\sum_{\alpha}\sum_{\beta}
\sum_{\gamma}\sum_{\delta}(\sigma_{zz})_{\alpha\beta,\gamma\delta}
C^{\dag}_{\alpha}C^{\dag}_{\beta}C_{\delta}C_{\gamma}\\
&=\frac{1}{2}[C^{\dag}_{+}C^{\dag}_{+}C_{-}C_{+}
-C^{\dag}_{+}C^{\dag}_{-}C_{+}C_{-}
+C^{\dag}_{-}C^{\dag}_{-}C_{-}C_{-}-
C^{\dag}_{-}C^{\dag}_{+}C_{+}C_{-}]
\end{split}
\end{equation}
When applied to a state $|+-)$ it yields
\begin{equation}
\begin{split}
\mathbf{\sigma_{zz}}|+-)&=\frac{1}{2}[C^{\dag}_{+}C^{\dag}_{+}C_{+}C_{+}|+-)-C^{\dag}_{+}C^{\dag}_{-}C_{-}C_{+}|+-)\\
&+C^{\dag}_{-}C^{\dag}_{-}C_{-}C_{-}|+-)-
C^{\dag}_{-}C^{\dag}_{+}C_{+}C_{-}]|+-)\\
&=\frac{1}{2}[-|+-)+|-+)]=-|+-)
\end{split}
\end{equation}
Thus, the mean value results
$(+-|\mathbf{\sigma_{zz}}|+-)=-(+-|+-)=-1$, which is the usual
result.

To obtain the correlation between components in two arbitrary
directions, say $\hat{z}$ and $\hat{n}$, we have to follow an
analogous procedure. First we write the operator $\sigma_{zn}$ that
acts over the state space of the two particles without
distinguishing them:
\begin{equation}\label{e:sigmazn}
\begin{split}
\mathbf{\sigma_{zn}}&=\frac{1}{2}[\cos\theta
C^{\dag}_{+}C^{\dag}_{+}C_{+}C_{+}+e^{-i\phi}\sin\theta
C^{\dag}_{+}C^{\dag}_{+}C_{-}C_{+}\\
&+e^{i\phi}\sin\theta C^{\dag}_{+}C^{\dag}_{-}C_{+}C_{+}-\cos\theta
C^{\dag}_{+}C^{\dag}_{-}C_{-}C_{+}\\
&+\cos\theta C^{\dag}_{-}C^{\dag}_{-}C_{-}C_{-}-e^{-i\phi}\sin\theta
C^{\dag}_{-}C^{\dag}_{-}C_{+}C_{-}\\
&-e^{-i\phi}\sin\theta C^{\dag}_{-}C^{\dag}_{+}C_{-}C_{-}-\cos\theta
C^{\dag}_{-}C^{\dag}_{+}C_{+}C_{-}]
\end{split}
\end{equation}
Applied to the state $|+-)$ it yields:
\begin{equation}
\begin{split}
\mathbf{\sigma_{zn}}|+-)&=\frac{1}{2}[e^{-i\phi}\sin\theta
C^{\dag}_{+}C^{\dag}_{+}|0)-\cos\theta
C^{\dag}_{+}C^{\dag}_{-}|0)\\
&+e^{-i\phi}\sin\theta C^{\dag}_{-}C^{\dag}_{-}|0)+\cos\theta
C^{\dag}_{-}C^{\dag}_{+}|0)]\\
&=\frac{1}{2}[-\cos\theta|+-)+\cos\theta|-+)]=\frac{1}{2}[-\cos\theta|+-)-\cos\theta|+-)]\\
&=-\cos\theta|+-)
\end{split}
\end{equation}
Thus, the mean value which gives the correlation results
$(+-|\mathbf{\sigma_{zn}}|+-)=-\cos\theta$, as it must be. It is
important to remark that the state $|+-)$ takes into account
indistinguishability and antisymmetry without `tricks', just because
it is constructed in $\mathbb{V}_{Q}$.

\section{Conclusions}\label{conclu}
We have argued that it is possible to construct a quantum
mechanical formalism for indistinguishable particles making use of
quasi-set theory  \Q\ to build a vector space $\mathbb{V}_{Q}$ that
resembles the Fock-space but without labeling quanta in any step. In
$\mathbb{V}_{Q}$, states refer only to occupation numbers and
permutations of quanta are unobservable. In this paper, we have
exemplified the use of the new formalism to evaluate the
correlations between the spin components of a two-fermions system,
explicitly showing that it is not necessary to first impose labels
to the particles and then masking the individuation by a
symmetrization postulate to obtain the usual results.

\end{document}